\documentclass[%
 reprint,
superscriptaddress,
 amsmath,amssymb,
 aps,
floatfix,
]{revtex4-1}
\usepackage{multirow}
\usepackage{graphicx}
\usepackage{dcolumn}
\usepackage{bm}
\usepackage{hyperref}
\DeclareUnicodeCharacter{200E}{ }

\begin{document}

\preprint{APS/123-QED}

\title{Cluster structure of 3$\alpha$+p states in $^{13}\mathrm{N}$}
		\author{J.~Bishop} 
		\affiliation{Cyclotron Institute, Texas A\&M University, College Station, TX 77843, USA}
        \affiliation{School of Physics and Astronomy, University of Birmingham, Edgbaston, Birmingham, B15 2TT, United Kingdom}
		\author{G.V.~Rogachev}
		\affiliation{Cyclotron Institute, Texas A\&M University, College Station, TX 77843, USA}
		\affiliation{Department of Physics \& Astronomy, Texas A\&M University, College Station, TX 77843, USA}
		\affiliation{Nuclear Solutions Institute, Texas A\&M University, College Station, TX 77843, USA}
		\author{S.~Ahn}
		\affiliation{Center for Exotic Nuclear Studies, Institute for Basic Science, 34126 Daejeon, Republic of Korea}
		\author{M.~Barbui}
		\affiliation{Cyclotron Institute, Texas A\&M University, College Station, TX 77843, USA}
		\author{S.M.~Cha}
		\affiliation{Center for Exotic Nuclear Studies, Institute for Basic Science, 34126 Daejeon, Republic of Korea}
		\author{E.~Harris}
		\affiliation{Cyclotron Institute, Texas A\&M University, College Station, TX 77843, USA}
		\affiliation{Department of Physics \& Astronomy, Texas A\&M University, College Station, TX 77843, USA}
		\author{C.~Hunt}
		\affiliation{Cyclotron Institute, Texas A\&M University, College Station, TX 77843, USA}
		\affiliation{Department of Physics \& Astronomy, Texas A\&M University, College Station, TX 77843, USA}
		\author{C.H.~Kim}
		\affiliation{Department of Physics, Sungkyunkwan University, Suwon 16419, Republic of Korea}
		\author{D.~Kim}
		\affiliation{Center for Exotic Nuclear Studies, Institute for Basic Science, 34126 Daejeon, Republic of Korea}
		\author{S.H.~Kim}
		\affiliation{Department of Physics, Sungkyunkwan University, Suwon 16419, Republic of Korea}
		\author{E.~Koshchiy}
		\affiliation{Cyclotron Institute, Texas A\&M University, College Station, TX 77843, USA}
		\author{Z.~Luo}
		\affiliation{Cyclotron Institute, Texas A\&M University, College Station, TX 77843, USA}
		\affiliation{Department of Physics \& Astronomy, Texas A\&M University, College Station, TX 77843, USA}
		\author{C.~Park}
		\affiliation{Center for Exotic Nuclear Studies, Institute for Basic Science, 34126 Daejeon, Republic of Korea}	
		\author{C.E.~Parker}
		\affiliation{Cyclotron Institute, Texas A\&M University, College Station, TX 77843, USA}		
		\author{E.C.~Pollacco}
		\affiliation{IRFU, CEA, Universit\'e Paris-Saclay, Gif-Sur-Yvette, France}		
		\author{B.T.~Roeder}
		\affiliation{Cyclotron Institute, Texas A\&M University, College Station, TX 77843, USA}
		\author{M.~Roosa}
		\affiliation{Cyclotron Institute, Texas A\&M University, College Station, TX 77843, USA}
		\affiliation{Department of Physics \& Astronomy, Texas A\&M University, College Station, TX 77843, USA}		
		\author{A. Saastamoinen}
		\affiliation{Cyclotron Institute, Texas A\&M University, College Station, TX 77843, USA}
		\author{D.P.~Scriven}
		\affiliation{Cyclotron Institute, Texas A\&M University, College Station, TX 77843, USA}
		\affiliation{Department of Physics \& Astronomy, Texas A\&M University, College Station, TX 77843, USA}
	
\email{jackbishop@tamu.edu}

\date{\today}

\begin{abstract}
\begin{description}
\item[Background]
Cluster states in $^{13}$N are extremely difficult to measure due to the unavailability of  $^{9}$B+$\alpha$ elastic scattering data.
\item[Purpose]
Using $\beta$-delayed charged-particle spectroscopy of $^{13}$O, clustered states in $^{13}$N can be populated and measured in the 3$\alpha$+p decay channel.
\item[Method]
One-at-a-time implantation/decay of $^{13}$O was performed with the Texas Active Target Time Projection Chamber (TexAT TPC). 149 $\beta 3\alpha p$ decay events were observed and the excitation function in $^{13}$N reconstructed.
\item[Results]
Four previously unknown $\alpha$-decaying excited states were observed in $^{13}$N at an excitation energy of 11.3 MeV, 12.4 MeV, 13.1 MeV and 13.7 MeV decaying via the 3$\alpha$+p channel.
\item[Conclusion]
 These states are seen to have a [$^{9}\mathrm{B}(\mathrm{g.s}) \bigotimes \alpha$/ $p+^{12}\mathrm{C}(0_{2}^{+})$], [$^{9}\mathrm{B}(\frac{1}{2}^{+}) \bigotimes \alpha$], [$^{9}\mathrm{B}(\frac{5}{2}^{+}) \bigotimes \alpha$] and [$^{9}\mathrm{B}(\frac{5}{2}^{+}) \bigotimes \alpha$] structure respectively. A previously-seen state at 11.8 MeV was also determined to have a [$p+^{12}\mathrm{C}(\mathrm{g.s.})$/ $p+^{12}\mathrm{C}(0_{2}^{+})$] structure. The overall magnitude of the clustering is not able to be extracted however due to the lack of a total width measurement. Clustered states in $^{13}$N (with unknown magnitude) seem to persist from the addition of a proton to the highly $\alpha$-clustered $^{12}$C. Evidence of the $\frac{1}{2}^{+}$ state in $^{9}$B was also seen to be populated by decays from $^{13}$N$^{\star}$.
\end{description}
\end{abstract}

\pacs{Valid PACS appear here}
\maketitle

\section{\label{sec:Introduction}Introduction}
The most well-known instance of $\alpha$ clustering in light nuclei is perhaps that of the Hoyle state in $^{12}$C \cite{HoyleFreer}. When additional protons or neutrons are added to this system, the propensity for clustering is of interest to study to understand the phenomenon of clustering. While clustering in $^{13}$C and $^{14}$C has been examined through resonant scattering - the unbound nature of $^{9}$B means that these data for $^{13}$N are not accessible. Instead, one must use different mechanisms to populate these highly-excited exotic states. Combined with the experimental difficulties of observing the 3$\alpha$+p decay of these states (as the characteristic clustered decay mode), a high-sensitivity experimental approach is required.\par
To probe these $\alpha$-clustered states in $^{13}$N, $\beta$-delayed charged-particle spectroscopy was used to populate states in $^{13}$N via $^{13}$O and decays to a final state of 3$\alpha$+p were then measured.
To achieve this, the Texas Active Target Time Projection Chamber (TexAT TPC) was used to perform one-at-a-time implantation and decay which has been demonstrated previously to have a very high sensitivity to rare decays due to the absence of background \cite{NIM,Hoyle,Efimov}. This paper provides more details on this approach and deeper insight into the states observed in a previous paper detailing the first observation of the $\beta 3\alpha p$ decay channel \cite{PRL}.

\section{\label{sec:setup}Experimental setup}
The experimental setup utilized for this experiment follows that successfully applied to studying the rare decay modes of near-threshold $\alpha$-clustered states in $^{12}$C \cite{NIM,Hoyle,Efimov} via the $\beta$-delayed 3$\alpha$-particle decay of $^{12}\mathrm{N}$ via $^{12}$C$^{\star}$. Instead, the $\beta$-delayed 3$\alpha p$-particle decay of $^{13}\mathrm{O}$ via $^{13}$N$^{\star}$ is studied here. \par
The K500 Cyclotron at Texas A\&M University was used to produce a beam of $^{14}$N which was incident on a $^{3}$He gas cell to produce a secondary beam of $^{13}$O via the $^{14}\mathrm{N}(^{3}\mathrm{He},xn)^{13}$O reaction. The $^{13}$O of interest was then selected using the MARS (Momentum Achromat Recoil Separator) \cite{MARS} with a typical intensity of 5 pps and an energy of 15.1 MeV/u. The beam was then degraded to 2 MeV/u using an aluminum foil in order to stop the beam inside of the sensitive area of the TexAT TPC \cite{TexATNIM} in 50 Torr of CO$_2$ gas.
The one-at-a-time $\beta$-delayed charged-particle spectroscopy technique requires the implantation of the $\beta$-decaying nucleus $^{13}$O into the active area of the TexAT TPC (with $t_\frac{1}{2}$ = 8.58 ms). The implanted ion then subsequently decays with the TPC volume being insensitive to the $\beta^+$ in the gas (and the subsequent $\beta^+$ from the decay of $^{13}$N for some events). When states in $^{13}$N are populated above the particle decay threshold (1.944 MeV for the p-threshold and 9.496 MeV for the $\alpha$-threshold), the daughter nucleus can undergo particle decay and the recoil products are measured inside the TPC. 
\par
\begin{figure}
\centerline{\includegraphics[width=0.5\textwidth]{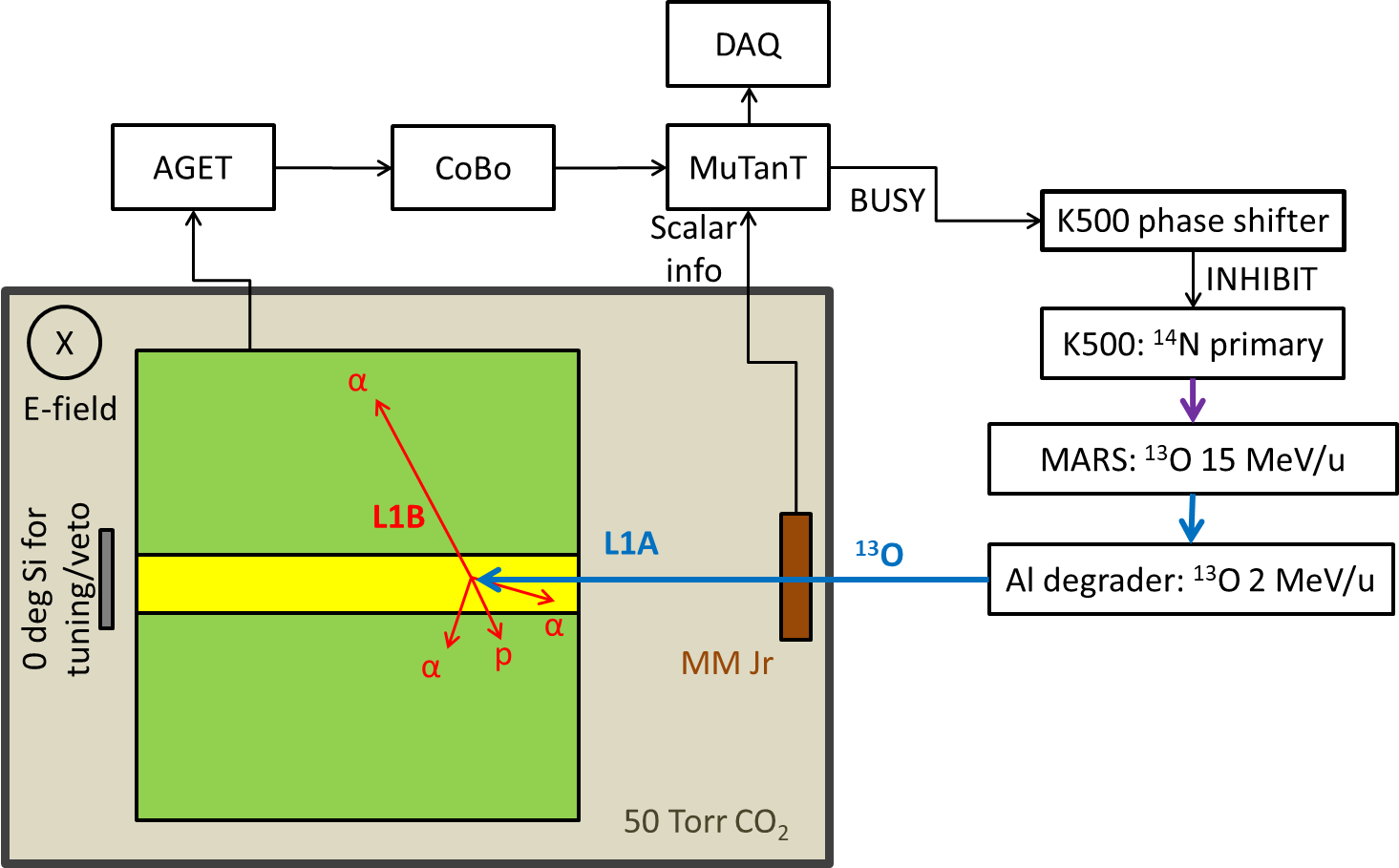}}
\caption{An overview of the experimental setup showing how the K500 Cyclotron phase shifter inhibits the $^{14}\mathrm{N}$ primary beam following an implantation event from the $^{13}\mathrm{O}$ secondary beam (L1A). Following this, for a fraction of events a subsequent decay (within 30 ms) of the $^{13}\mathrm{N} \rightarrow 3\alpha+p$ (or more-likely, a single proton) provides a second L1B (decay) trigger and the decay products from can be reconstructed inside the TPC. A silicon detector at zero degrees was used for tuning the radioactive beam and providing a veto signal for beam events that do not stop inside of the TexAT active region. \label{fig:setup}}
\end{figure}

As the ion is incident into the TPC active volume, the implantation beam track triggers the GET electronics setup \cite{GET} which sends a 30-ms-long busy signal to the K500 phase shifter which blocks the transmission of the primary $^{14}$N through the cyclotron and therefore prevents any subsequent implantations. During the 30 ms busy signal, the data acquisition then awaits an additional event corresponding to the decay of $^{13}$N$^{\star}$ through the proton or $\alpha$ decay channels. For the majority of events, this second trigger is not generated as the $^{13}$O decays to states in $^{13}$N that are particle bound. After either 30 ms has elapsed or a decay event has registered, the data acquisition is ready for a new implantation event and the K500 phase shifter signal is disabled. This setup is known as the `2p-mode' in the GET system Mutant module and allows for the beam implant and decay events to be correlated. The implantation from the beam is referred to as the L1A trigger and requires 10 channels in TexAT to be above threshold. The decay event is known as the L1B trigger and requires only a multiplicity of 1 to allow for very low-energy recoils to trigger the data acquisition. The time between the L1A and L1B event (known as the d2p time) is also recorded in the GET system and corresponds to the decay time of the $\beta$-delaying particle. Partial events where an L1A trigger was not followed by an L1B trigger were also recorded for normalization and beam characterization. An overview of this experimental setup is shown in Fig.~\ref{fig:setup}.

\section{Methodology}
\begin{figure}
\centerline{\includegraphics[width=0.5\textwidth]{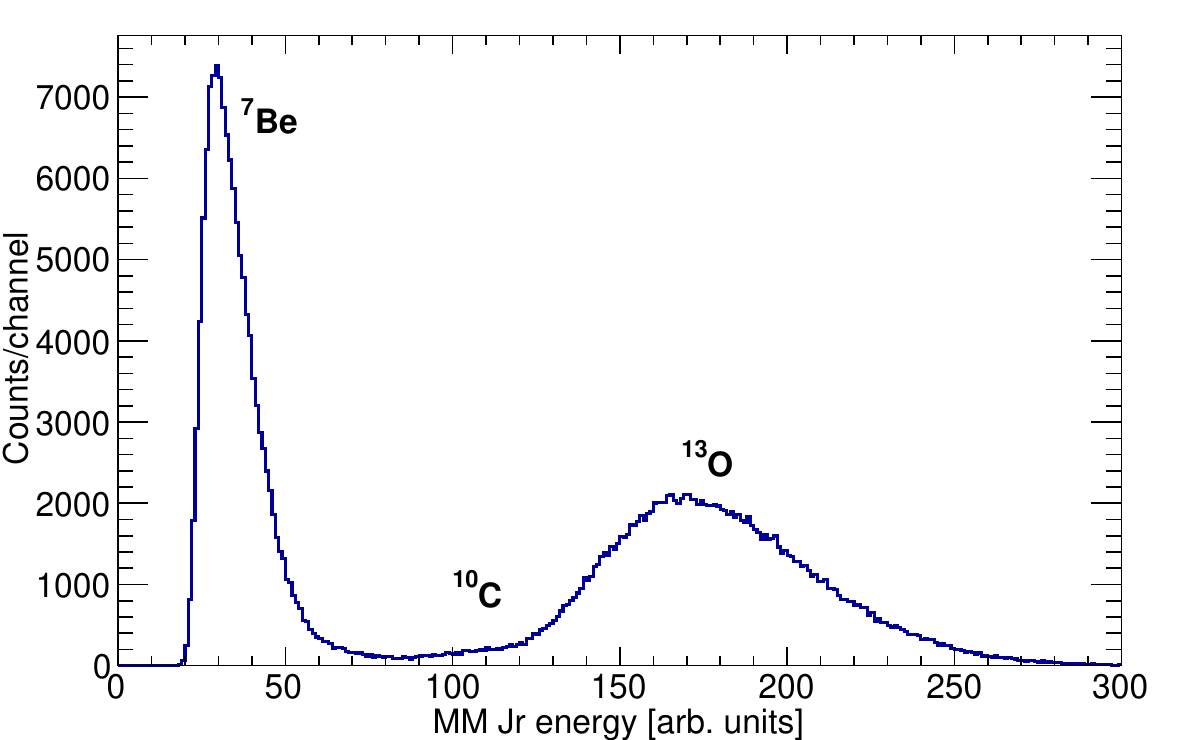}}
\caption{Energy deposition (dE) within the Micromegas Jr detector to differentiate the beam species. The contaminants to the $^{13}\mathrm{O}$ were dominated by $^{7}\mathrm{Be}$ and they total $\approx$ 28\% of the total beam intensity of 5 pps.\label{fig:MMJrE}}
\end{figure}

To select events of interest, the full L1A (implant) + L1B (decay) events were selected where the time between the two was between 1 and 30 ms (with small times omitted to remove double trigger events due to sudden beam-induced noise). In addition to the secondary $^{13}$O of interest, there were also some other beam contaminants. Therefore, to ensure the implanted ion corresponded to $^{13}\mathrm{O}$, the energy deposited by the beam implant event in the Micromegas ``Jr" (MM Jr) beam tracker \cite{MMJr} at the entrance to the TexAT chamber was recorded. The beam contaminants were $^{7}\mathrm{Be}$ and $^{10}\mathrm{C}$, dominated by $^{7}\mathrm{Be}$ at $\approx$ 28\% of the beam intensity. The energy spectrum from the Micromegas Jr is shown in Fig.~\ref{fig:MMJrE}.\par
\begin{figure}
\centerline{\includegraphics[width=0.5\textwidth]{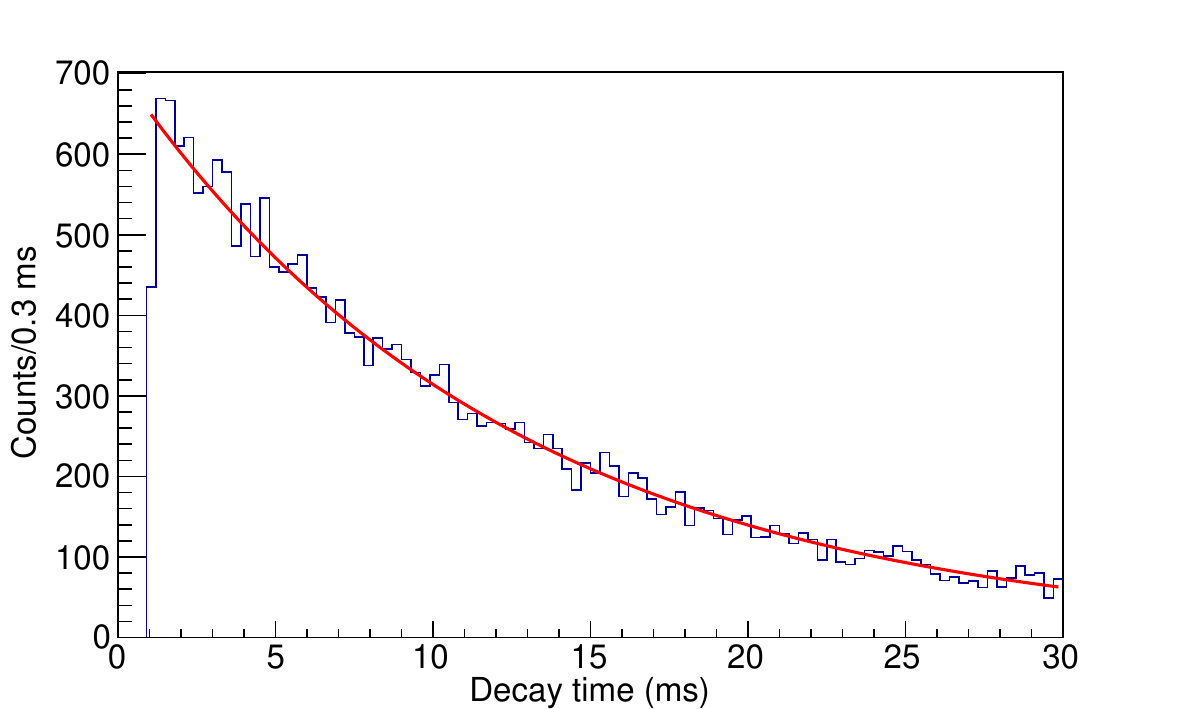}}
\caption{Decay time between the implant of $^{13}\mathrm{O}$ and decay overlaid with a background free exponential fit yielding a value of $t_{1/2}$=8.55$\pm$0.09 (stat.) ms compared to the adopted value of 8.58$\pm$0.05 ms \cite{O13t}. $\tilde{\chi}^2$ = 1.26 from log-likelihood minimization.\label{fig:d2p}}
\end{figure}
Following a selection of $^{13}\mathrm{O}$ implants, the stopping position of the beam was evaluated event-by-event and events where the beam stopped inside the active area of the Micromegas were taken, events where a signal was observed in the zero-degree silicon detector were automatically vetoed as signifying the beam did not stop in the gas. Because of the proximity of some events to the edge of the active region, only events which were more than 31.5 mm from the edge of the Micromegas were taken for further analysis. Due to the large energy degradation required to stop inside TexAT, the spread in the $^{13}\mathrm{O}$ stopping position was 67.5 mm.\par
To ensure the implant and decay events are correlated, the 2D distance (in the plane of the Micromegas) between the stopping position of the implanted beam and the decay vertex (identified by the highest energy deposition inside the active volume) is then evaluated. Following previous results \cite{NIM}, the separation for correlated events is given by:
\begin{eqnarray}
f(r) = re^{-\frac{(r-r_{0})^2}{2\sigma^2}},
\end{eqnarray}
with $\sigma$ = 4.9 mm for these data which agrees well with a predicted diffusion value of $\approx 4$ mm from the ideal gas law. A selection of events with a displacement of $<5$ mm were taken as correlated. To determine the purity of this selection, the time between the implantation and decay for these events was accumulated (the d2p time). A background-free exponential fit yielded a half-life value of 8.55$\pm$0.09 (stat.) ms compared to the adopted value of 8.58$\pm$0.05 ms ($\tilde{\chi}^2$ = 1.26) which is shown in Fig.~\ref{fig:d2p}. The excellent agreement without the need for any background demonstrates the purity of the selection of $^{13}\mathrm{O}$ in the system.\par
Tracks from these events were fit with a single track segment using a randomly-sampled $\chi$-squared minimization algorithm (more detail on this fitting technique is covered in Sec.~\ref{sec:3ap}). For those tracks for which the reduced $\chi$-squared was good, these events were identified as single proton events and are discussed in Sec.~\ref{sec:proton}. For those which gave a poor reduced $\chi$-squared, these events were fit with four track segments as candidate 3$\alpha$+p events using randomly-sampled $\chi$-squared minimization and were visually inspected to evaluate the quality of fit and also identify any events which were unable to be fit with the algorithm (given the complexity of the fit required). These 3$\alpha$+p events are discussed in Sec.~\ref{sec:3ap}.
\section{Single proton events\label{sec:proton}}
Due to the low gas pressure used, the majority of proton event tracks escape the TPC active volume. In order to evaluate the state populated in $^{13}\mathrm{N}^{\star}$, the recoiling $^{12}\mathrm{C}$ was therefore used instead which has 1/12$^{th}$ of the proton energy and a much higher dE/dx therefore stops safely inside the active volume. \par
\begin{figure}
\centerline{\includegraphics[width=0.5\textwidth]{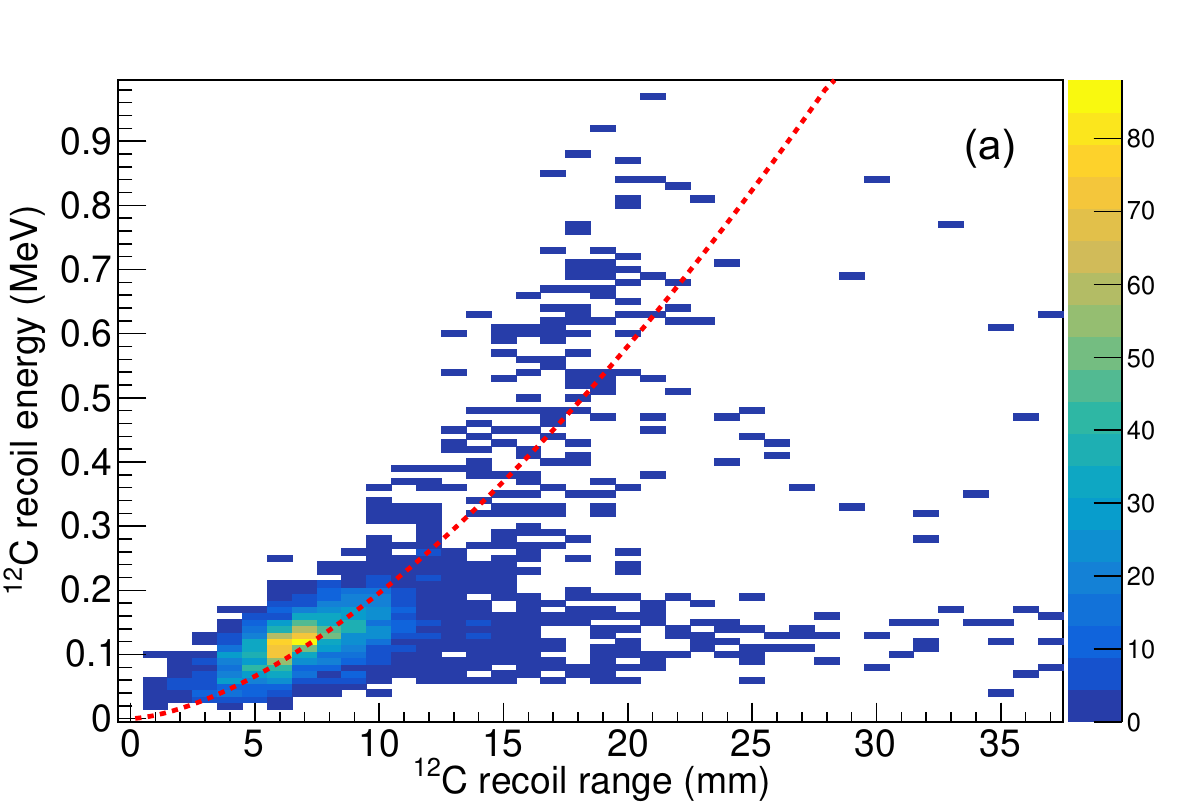}}
\centerline{\includegraphics[width=0.5\textwidth]{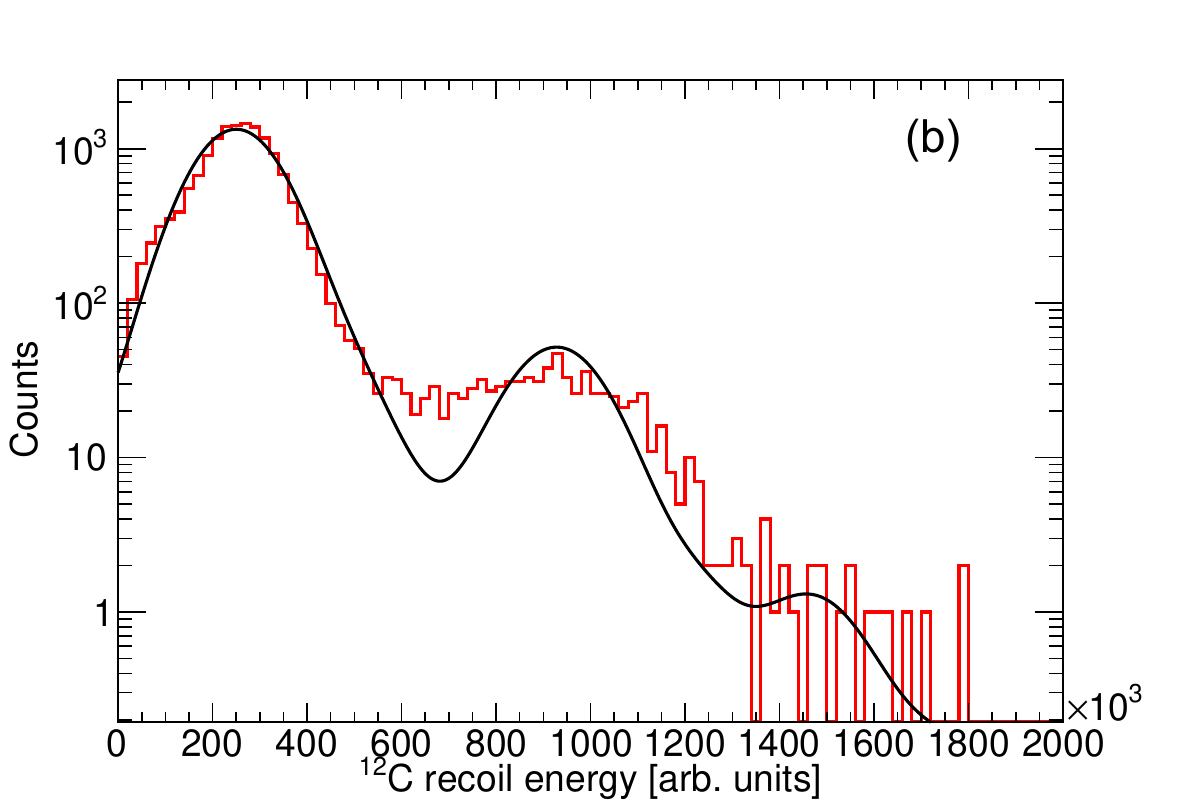}}
\caption{Selection of ($^{13}\mathrm{N},p_{0,1})$ events. Top: Experimental data of $^{12}$C recoil range in TPC versus energy. The expected result from TRIM \cite{SRIM} is overlaid as a dotted red line. Bottom: Proton energy spectrum from these data obtained from the $^{12}\mathrm{C}$ recoil (red histogram). The expected yield using previously-obtained branching ratios \cite{Knudsen} is overlaid as a black solid line after being convoluted with a Gaussian profile to best replicate the data.\label{fig:13N}}
\end{figure}
In order to calculate the $^{12}\mathrm{C}$ recoil energy, the decay event is fit with a single line segment defined by two points: the end point of the proton track and the end point of the $^{12}\mathrm{C}$ track. The decay vertex then dissects this line segment and the energy in the point cloud that is on the recoil side of the fit was also added to the $^{12}\mathrm{C}$ recoil energy. Fig.~\ref{fig:13N}a shows the relationship between the recoil energy and range from the track overlaid with the expected TRIM result which allowed for an additional cut on the data. The events that lie outside of this cut constitute events where the separation between the proton and $^{12}\mathrm{C}$ recoil could not be accurately attained. The number of events lost through this cut is $<4\%$ which has little effect on the branching ratios but removes all background at higher excitation energies. Given the typical energy of the recoil is from 0.12 to 1.1 MeV (corresponding to track lengths of 6.7 to 30.5 mm), the resolution of the energy spectrum using this technique is poor and the proton energies and relative intensities from previous studies \cite{Knudsen} are used to demonstrate the compatibility of our result with those previously observed. The excitation function obtained is shown in Fig.~\ref{fig:13N}b with the expected yield overlaid from previous studies \cite{Knudsen} convoluted with a Gaussian response. Any deviation can be attributed to the dynamics of the gas avalanche that is better characterized as a Breit-Wigner convoluted with a Gamma distribution (representing the gas gain dynamics), convoluted with a Gaussian (representing the resolution associated with noise and difficulties associated with fitting the $^{12}$C recoil track). This also confirms the implanted number of $^{13}\mathrm{O}$ that make it through our cuts is 1.90$\times 10^{5}$ from 1.86$\times 10^{4}$ proton decay events. An independent branching ratio measurement from the number of implants was not reliable during this experiment due to a sizeable noise contribution which adversely affected the L1A/L1B ratio (corresponding to the branching ratio to particle-unbound states) but was necessary to ensure 100\% trigger efficiency on genuine proton events.
\section{3$\alpha$+proton events\label{sec:3ap}}
A total of 149 3$\alpha$+p events were identified. Due to the size of the TPC and limitations on reconstruction in parts of the TexAT TPC, only 102 out of 149 of these events could be fully-reconstructed. These events that are lost are almost exclusively from $\alpha$-decay as this produces a high-energy $\alpha$-particle that may escape. The efficiency for the $\alpha_0$ decay starts to deviate from 100\% at $E_{x}$ = 10 MeV, slowly drops to around 60\% at $E_{x}$ = 14 MeV. The efficiency for $\alpha_1$ and $\alpha_3$ are less affected and only decrease to 70\% at $E_{x}$ = 14 MeV. For events that proton decay to the Hoyle state, the majority of the energy is taken by the proton which is not required for reconstruction. Corrections to the yields obtained during this work are made to account for this effect in Table~\ref{tab:states}.\par
\begin{figure}
\centerline{\includegraphics[width=0.5\textwidth]{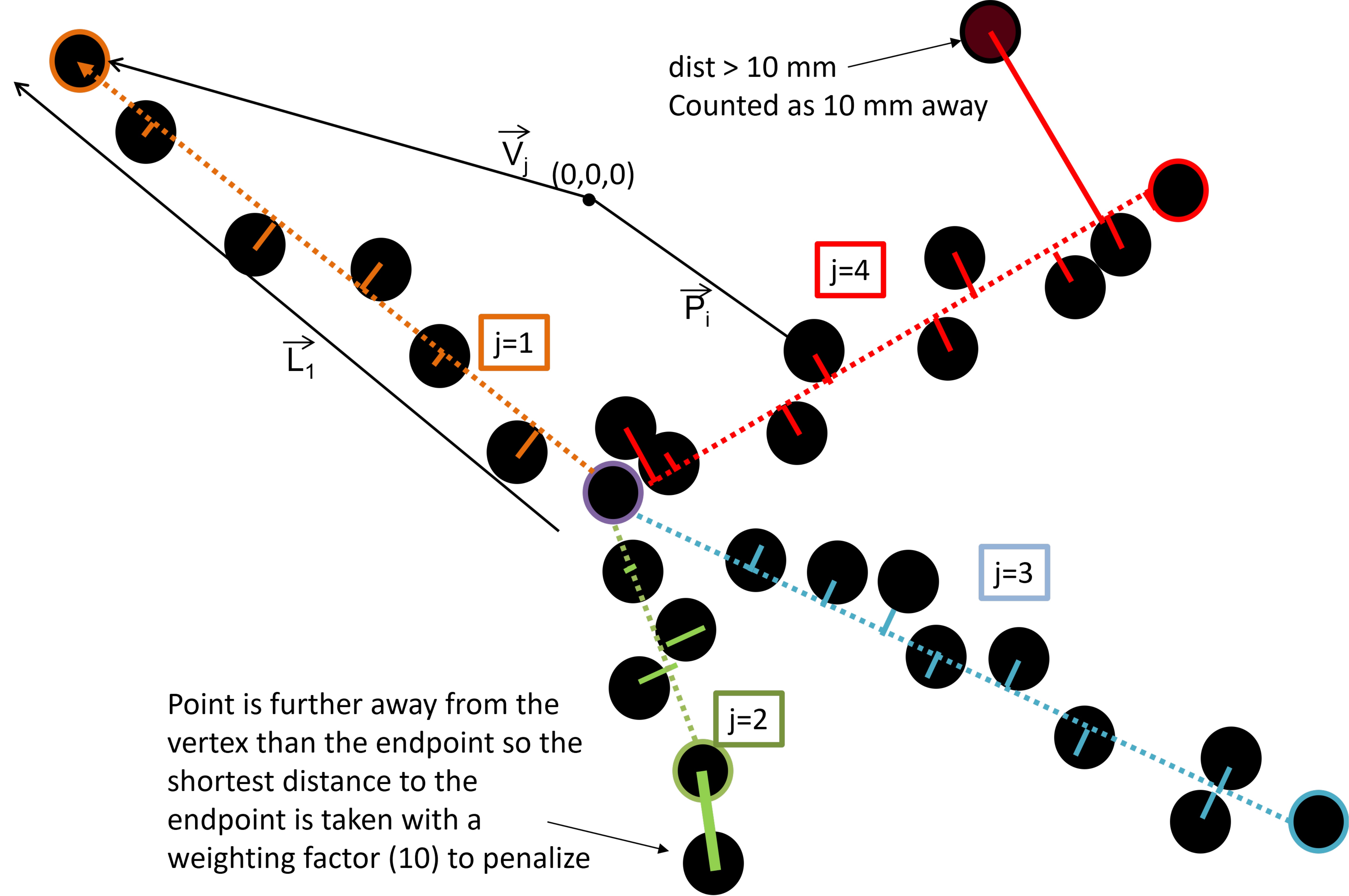}}
\caption{Schematic showing the basis of the RANSChiSM fit by selection of five points to parameterize the four-track fit. Any points in the pointcloud more than 10 mm from the nearest track line are counted as if they are 10 mm away to reduce the influence of outliers.\label{fig:RANSAC3ap}}
\end{figure}

In order to accurately fit the four-track events, a highly-robust fitting technique is required. In order to achieve this, the ansatz for the reaction vertex is identified by the stopping position of the beam and then the point cloud for the decay events is fit with 15 parameters: the decay vertex and the endpoint of each of the four tracks (with each being a 3D vector). Due to the highly complex nature of the fit with many free parameters and a `noisy' fitting space, a modified version of RANSAC \cite{RANSAC} was used that has been successfully employed in several other TexAT experiments whereby the four parameters are randomly selected from the set of points in the pointcloud (as per RANSAC) and then a goodness of fit is evaluated by the sum of the distance-squared for all points to the nearest line defined by the vertex to each of the three endpoints. It is better characterized as RANdomly-Sampled Chi-Squared Minimization, referred to here as RANSChiSM. To reduce the influence of outliers, where the distance to all three lines exceeded 10 mm, the distance was saturated to be 10 mm. This functional form is given by:
\begin{eqnarray}
\tilde{\chi}^2 = \frac{\Sigma_i^{N_\texttt{points}}}{N_{\texttt{points}}} \left(\texttt{min}_{j=1,2,3,4} \left[\frac{|(\vec{P_i}-\vec{V_j}) \times \vec{L_j}|}{|\vec{L_j}|},10\right]^2 \right),
\end{eqnarray}
where $\vec{P_i}$ is the i$^{th}$ point in the pointcloud, $\vec{V_j}$ is the vector for the endpoint of the $j^{th}$ track, and $\vec{L_j}$ is the vector from the decay vertex to the endpoint of one of the three tracks. The vector product here merely calculates the shortest 3D distance of the point $\vec{p_i}$ to the line $\vec{L_j}$ To ensure the point that is chosen is at the end of the track, if the distance between the point and the vertex exceeds the distance between the vertex and the end of the arm, then the distance between the point and the end of the arm is taken instead, weighted by a factor of 10. This was seen to be very effective at ensuring the proper selection of the end of the tracks which is necessary for the conversion between range in the gas and the energy. To reduce the number of random samples of endpoints and the decay vertex, one endpoint was defined as that which had the largest distance from the decay vertex and a weighting scheme was used where two of the remaining three endpoints were randomly sampled with a weight given by the distance of the point in the pointcloud to the randomly-chosen decay vertex squared. Similarly, only points within 5 mm of the known beam stopping point were allowed to be selected for the decay vertex point. A schematic of this fitting is shown in Fig.~\ref{fig:RANSAC3ap}. For the $\beta 3 \alpha p$ events, 20,000 iterations were sufficient to converge on the best fit. Each decay was examined manually and those that were not perfectly fitted were reanalyzed and, if needed, the parameters for certain tracks were constrained manually and the fit reevaluated. For some extremely short tracks, this was necessary due to the presence of many `fake' minima in the chi-squared phase space.

\begin{figure}
\centerline{\includegraphics[width=0.5\textwidth]{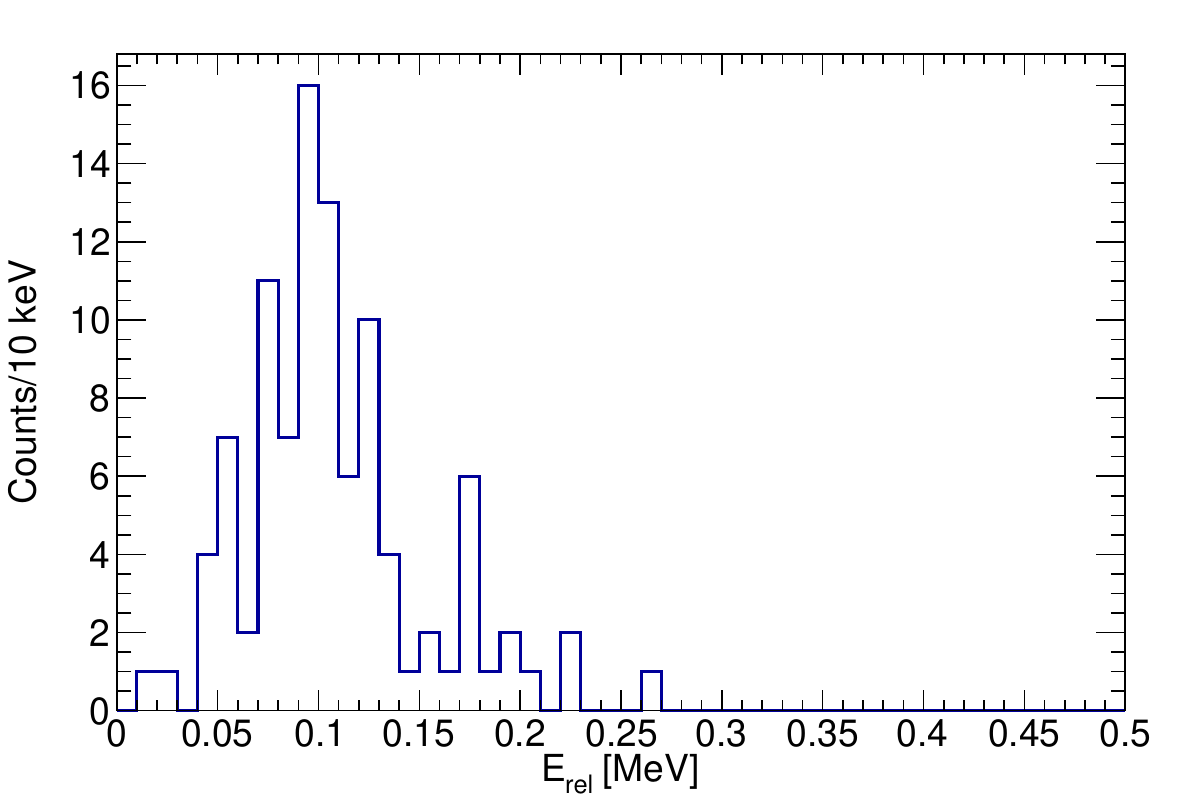}}
\caption{Relative energy spectrum for pairs of $\alpha$-particles (taking the smallest relative energy) showing the $^{8}\mathrm{Be}$(g.s) energy of 92 keV is well-reproduced in our data.\label{fig:8Be}}
\end{figure}

\begin{figure}
\centerline{\includegraphics[width=0.5\textwidth]{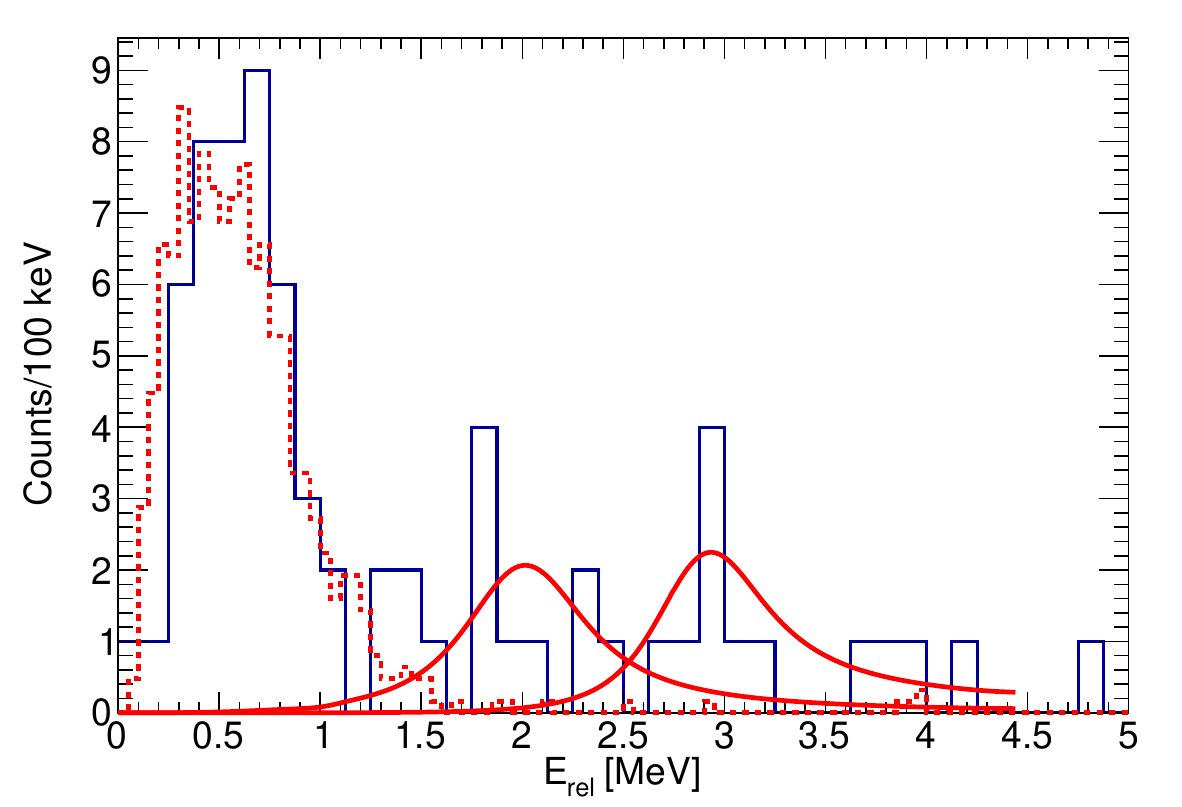}}
\caption{For events that do not decay via the Hoyle state, the relative energy spectrum is shown here which is generated by selecting the two $\alpha$-particles that produce the $^{8}\mathrm{Be}$(g.s) and then reconstructing the $^{9}\mathrm{B}$ relative energy with the proton. Overlaid in dashed red are simulated data for the ground state contribution and in solid red are the $\frac{1}{2}^{+}$ and $\frac{5}{2}^{+}$ states from single channel R-Matrix calculations convoluted with a Gaussian with $\sigma$ = 0.23 MeV. The $\frac{1}{2}^{+}$ parameters are those obtained by Wheldon \cite{Wheldon} which show excellent agreement.\label{fig:9B}}
\end{figure}
\begin{figure}
\centerline{\includegraphics[width=0.5\textwidth]{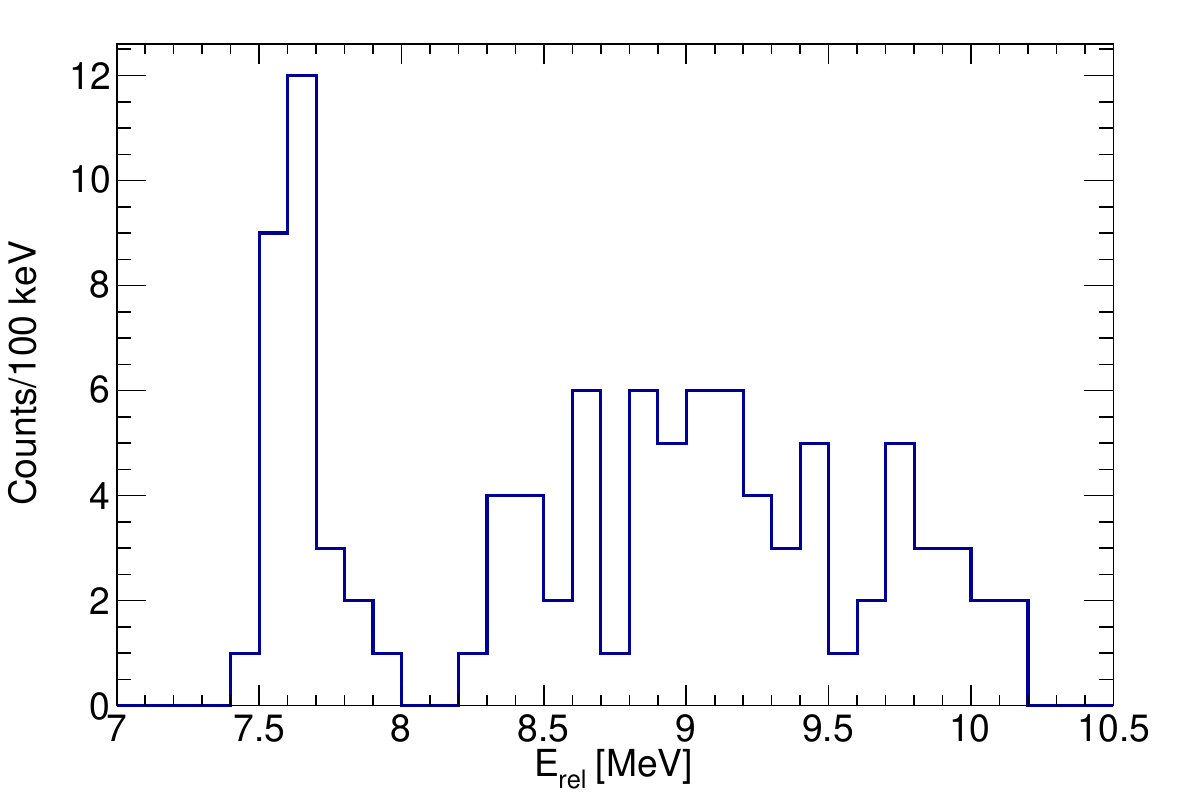}}
\caption{Invariant mass spectrum for $^{12}\mathrm{C}$ from 3$\alpha$-particles. A peak at 7.65 MeV is seen, well reproducing the Hoyle state energy and a broad peak is seen at higher excitation energies which correspond to events that decay via $^{9}\mathrm{B}+\alpha$.\label{fig:12C}}
\end{figure}
\begin{figure}
\centerline{\includegraphics[width=0.5\textwidth]{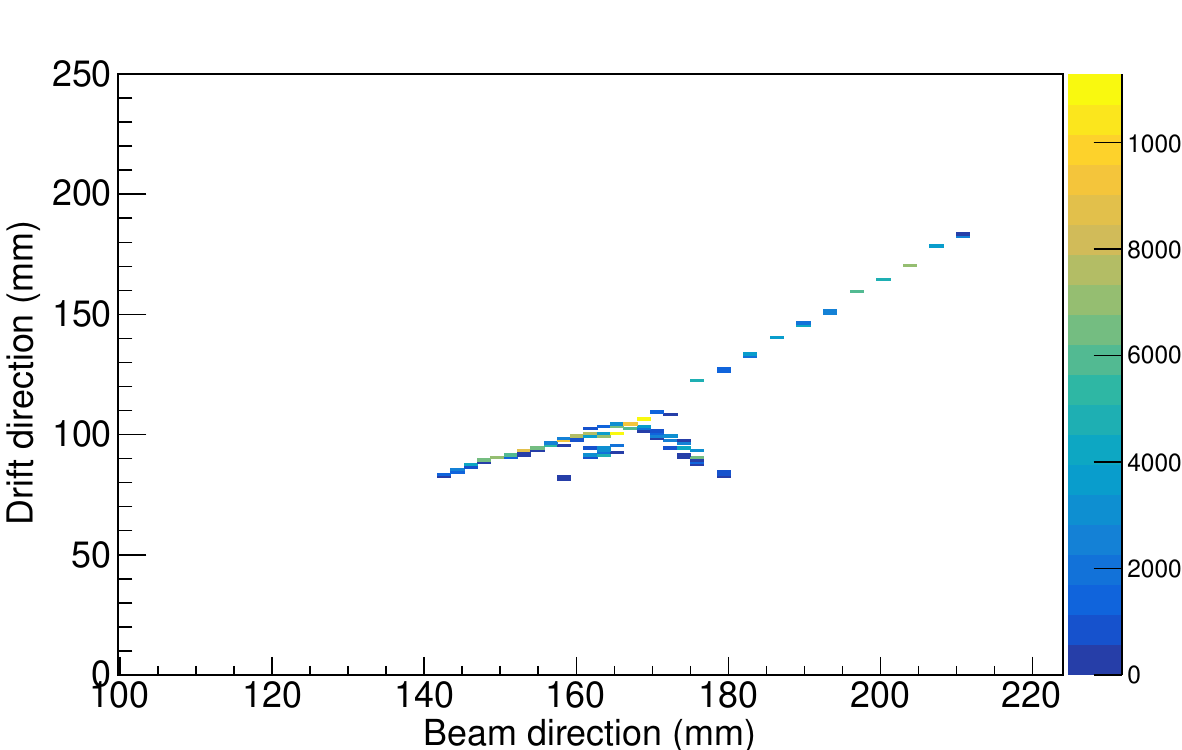}}
\caption{An example $p_2$ ($^{12}\mathrm{C}(0_{2}^{+})+p$) event where the energy deposition as a function of distance in the TPC is shown projected in 2D.\label{fig:HoyleEvent}}
\end{figure}
\begin{figure}
\centerline{\includegraphics[width=0.5\textwidth]{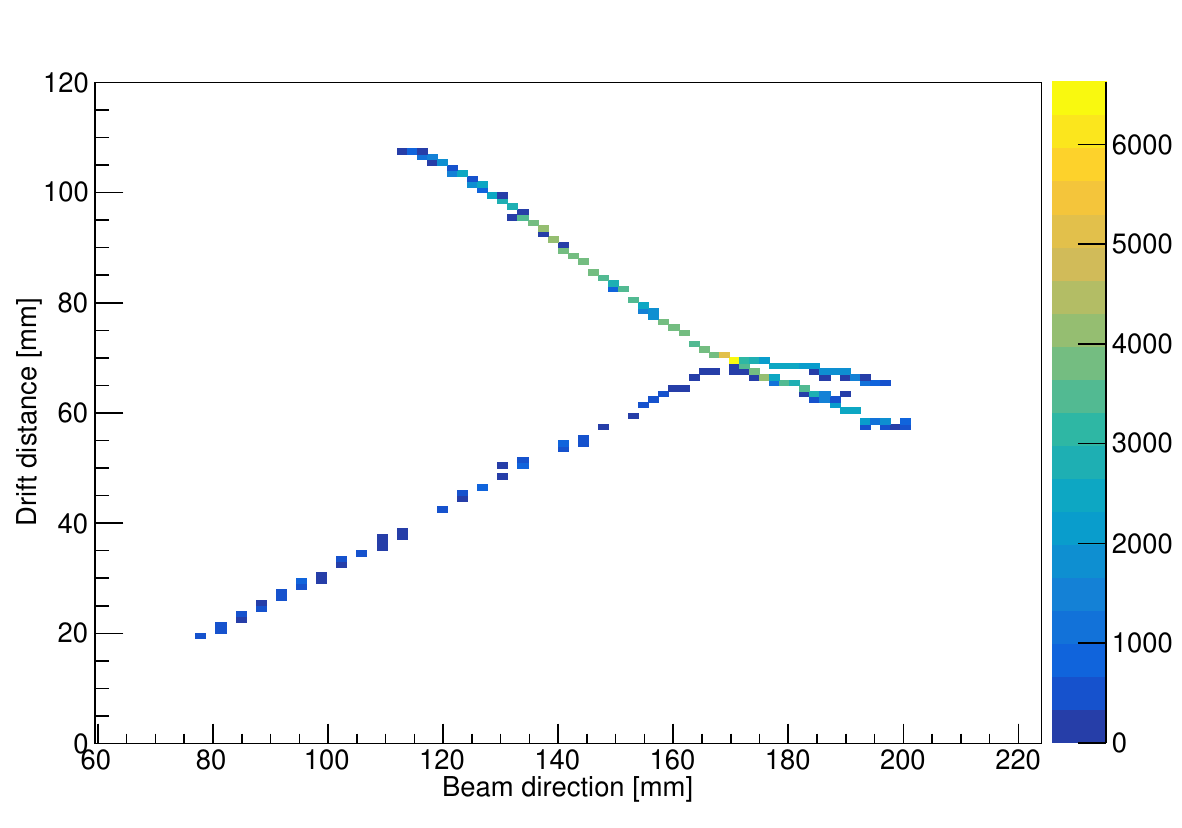}}
\caption{An example $\alpha_0$ ($^{9}\mathrm{B}(\mathrm{g.s.})+\alpha$) event where the energy deposition as a function of distance in the TPC is shown projected in 2D. The track going downwards can be identified as the proton by its lower energy deposition. Also shown in \cite{PRL}. \label{fig:AlphaEvent}}
\end{figure}

In order to identify the parent state in $^{13}\mathrm{N}^{\star}$, the lowest energy deposition arm was identified as the proton track and the momentum of the 3 $\alpha$-particles was determined by the length and direction of the track in the gas. As the proton almost always escapes the TPC sensitive volume, the proton momentum is reconstructed from momentum conservation. The decay energy is then the sum of the three $\alpha$-particles and proton energy. From here, the $^{8}\mathrm{Be}$ (Fig.~\ref{fig:8Be}), $^{9}\mathrm{B}$ (Fig.~\ref{fig:9B}) relative energies and $^{12}\mathrm{C}$ (Fig.~\ref{fig:12C}) excitation energies were determined from the invariant mass. This allowed for a selection of events which proceeded to decay via p+$^{12}\mathrm{C}(0_{2}^{+})$ [$p_{2}$], $\alpha$+$^{9}\mathrm{B}$(g.s) [$\alpha_{0}$], $\alpha$+$^{9}\mathrm{B}(\frac{1}{2}^{+})$ [$\alpha_{1}$] and $\alpha$+$^{9}\mathrm{B}(\frac{5}{2}^{+})$ [$\alpha_{3}$]. An example $p_2$ event is shown in Fig.~\ref{fig:HoyleEvent} and an example $\alpha_0$ event is shown in Fig.~\ref{fig:AlphaEvent}. It is remarkable to see in Fig.~\ref{fig:9B} evidence of strength in $^{9}\mathrm{B}$ between 1 and 2.4 MeV that cannot be explained without the long-sought after $\frac{1}{2}^{+}$ state in $^{9}\mathrm{B}$ that is the mirror of the well-studied $^{9}\mathrm{Be}$ $\frac{1}{2}^{+}$. The ground state contribution (from Monte Carlo simulations) and the higher-lying states from a single-channel Breit-Wigner convoluted with a Gaussian ($\sigma$=0.23 MeV) were fitted to the spectrum. Attempting to fit the spectrum with and without the $\frac{1}{2}^{+}$ contribution shows the probability such a peak can occur by chance is 0.01\% demonstrating this contribution can only occur due to a contribution from the $\frac{1}{2}^{+}$ due to the absence of any background. The $\frac{1}{2}^{+}$ state in $^{9}\mathrm{B}$ was selected by taking an excitation energy of between 1.4 and 2.4 MeV in $^{9}\mathrm{B}$ (following the centroid and width as observed via $^{9}\mathrm{Be}(^{3}\mathrm{He},t)$ \cite{Wheldon} which is consistent with our current results) and the $\frac{5}{2}^{+}$ was taken as having an excitation energy of above 2.4 MeV. A more recent experimental study has suggested the possibility that discrepancies in experimental results are the result of a doublet of $\frac{1}{2}^{+}$ states \cite{Doublet}. Any contribution from the relatively-narrow 2.345 MeV $\frac{5}{2}^{-}$ is not present in the presented plots as this state decays almost exclusively via $^{5}\mathrm{Li}$ and therefore would not correspond to a peak in the $^{8}$Be spectrum. There were only 3 events associated with this decay to $^{5}\mathrm{Li}$ hence the statistics were insufficient to incorporate into the analysis. 

\par
\begin{figure}
\centerline{\includegraphics[width=0.5\textwidth]{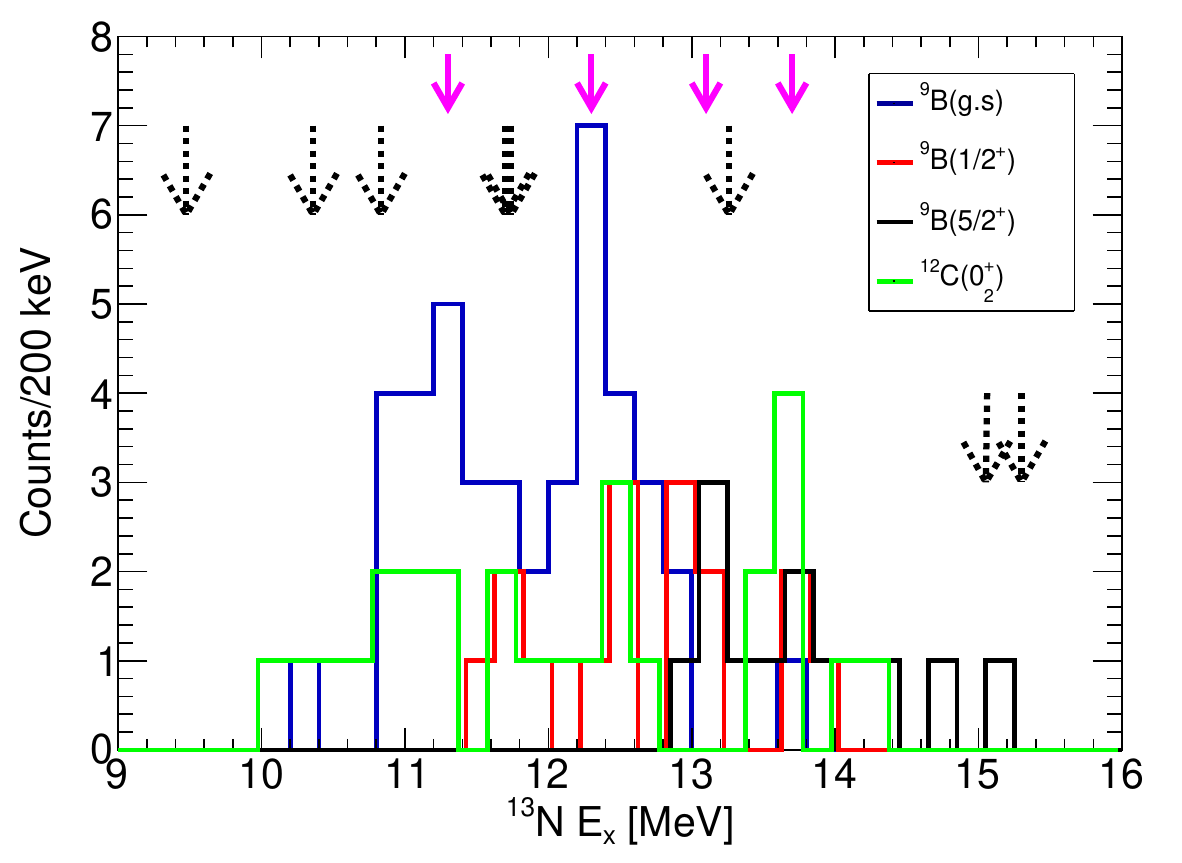}}
\caption{Excitation spectrum in $^{13}\mathrm{N}$ for 3$\alpha+p$ separated by channels. Dashed vertical arrows show previously-known states populated by $\beta$-decay in black and new states observed are shown by solid magenta arrows. Also shown in \cite{PRL}. \label{fig:3ap}}
\end{figure}
\begin{figure}
\centerline{\includegraphics[width=0.5\textwidth]{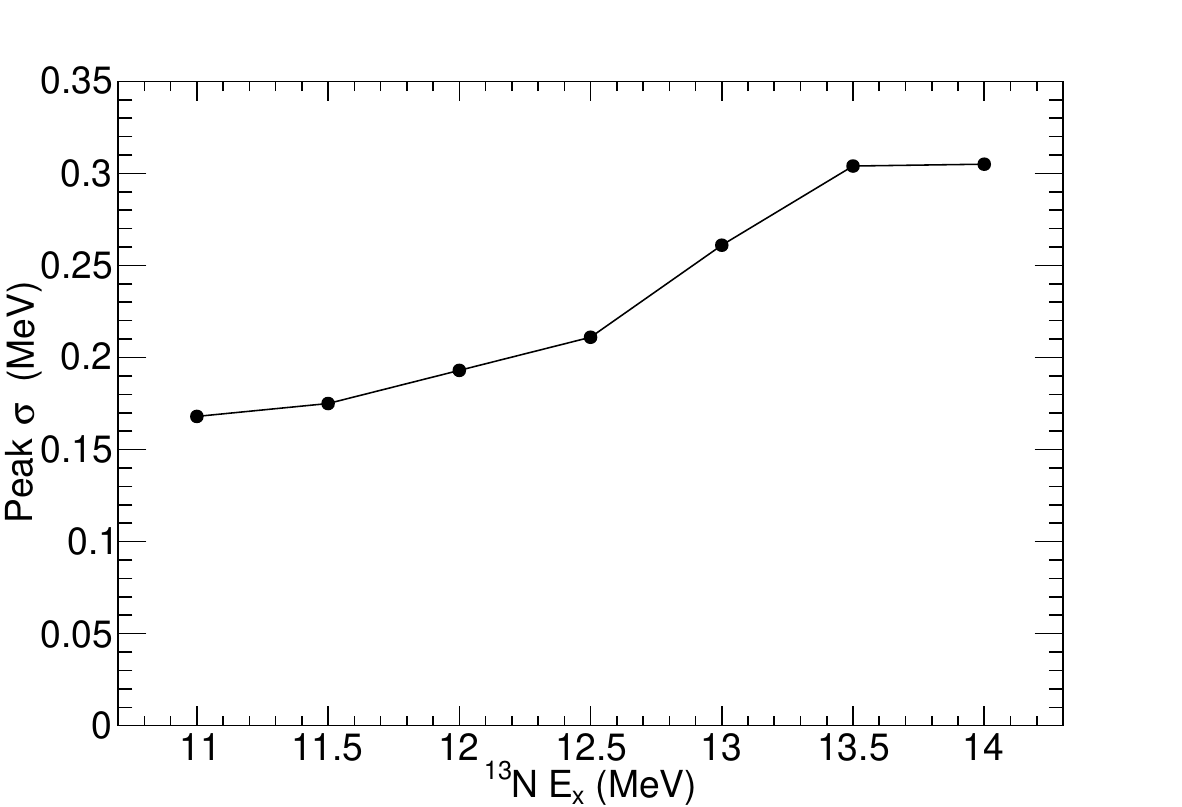}}
\caption{Variation of experimental resolution from a GEANT4 simulation of infinitely narrow states decaying via $^{9}$B(g.s)+$\alpha$ as a function of excitation energy. The statistical error bars are not shown as the error for each point is dominated by systematic errors via the stopping power uncertainties for $\alpha$-particles and may be as large as 0.1 MeV (towards larger peak $\sigma$ values).\label{fig:resolution}}
\end{figure}

Following the channel selection, the excitation energy in $^{13}\mathrm{N}$ was calculated and is shown in Fig.~\ref{fig:3ap}. Despite low statistics, a number of states can be seen and will be discussed individually. A summary of the properties of these states observed is then shown in Table~\ref{tab:states}. A GEANT4 simulation was performed to test the variation in experimental resolution as a function of excitation energy for the $\alpha_0$ channel which, as shown in Fig.~\ref{fig:resolution}, is typically around $\sigma$ = 200 keV. The $p_2$ channel resolution is almost entirely dominated by discrepancies between the expected and real stopping powers for the $\alpha$-particles and therefore cannot be accurately determined and is extremely sensitive to small changes. For all excitation energies, it is realistically greater than $\sigma$ = 160 keV however.


\par
\begin{table*}[!ht]
\setlength\extrarowheight{2.0pt}
    \caption{\label{tab:states}States in $^{13}\mathrm{N}$ observed in the current work with a preferred (but not definite) spin-parity assignment along with the decay properties of the states, calculated from the yields for the state in several different decay channels. The fractional reduced width is also efficiency corrected.}
    \centering
    \begin{tabular}{|l||l||l|l|l||l|l|l||l|l|l|l|l|l|}
    \hline
        \multicolumn{2}{|c|}{State} & \multicolumn{6}{|c|}{Counts} &  \multicolumn{6}{|c|}{Efficiency-corrected $\bar{\gamma}^2$} \\ \hline
        $E_{x}$ & $J^{\pi}$ & $\alpha_0$ & $\alpha_1$ & $\alpha_3$ & $p_0$ \cite{Knudsen} & $p_1$\cite{Knudsen} & $p_2$ & $\alpha_0$ & $\alpha_1$ & $\alpha_3$ & $p_0$ & $p_1$ & $p_2$ \\ \hline
        11.3(1) & 3/2- & $18(4.4)$ & 0 & 0 & $6(2.6)$ & $<3$ & $7(2.8)$ & 67(21)\% & 0\% & 0\% & 4(2)\% & $<$1\% & 29(13)\% \\ \hline
        11.8(1) & 3/2- & $<1.8$ & 0 & 0 & $28(14)$ & $<4$ & $4(2.2)$ & $<$12\% & 0\% & 0\% & 50(30)\% & 0\% & 38(25)\% \\ \hline
        12.4(1) & 3/2- & $22(4.8)$ & $4(2.2)$ & 0 & $<3$ & $<10$ & $5(2.5)$ & 6(2)\% & 88(49)\% & 0\% & $<$0.1\% & $<$2\% & 2(1)\% \\ \hline
        \multirow{2}{*}{13.1} & 1/2- &  \multirow{2}{*}{0} & \multirow{2}{*}{$3(2)$} & \multirow{2}{*}{$5(2.5)$} & \multirow{2}{*}{$21(6)$} & \multirow{2}{*}{$<10$} & \multirow{2}{*}{0} & 0\% & 1(1)\% & 98(48)\%\footnote{ Here the $\alpha_3$ channel is assumed to be through the $J^{\pi}=\frac{1}{2}^{-}$ channel in $^{9}$B rather than the $J^{\pi}=\frac{5}{2}^{+}$ state.} & 0\% & $<$0.4\% & 0\% \\ 
         & 5/2- &  &  & & & & & 0\% & 10(10)\% & 89(44)\% & 0.7(0.2)\% & $<$0.2\% & 0\% \\ \hline
        13.7(1) & 3/2- & $1(1.4)$ & $3(2)$ & $4(2.2)$ & $<3$ & $<10$ & $6(2.7)$ & 1(1)\% & 8(8)\% & 75(42)\% & $<$0.5\% & $<$7\% & 8(3)\% \\ \hline
    \end{tabular}
\end{table*}
\subsection{11.3 MeV state}
The first peak in the spectrum corresponds to an excitation energy of 11.3 MeV in $^{13}\mathrm{N}$. The strength is almost entirely dominated by the $^{9}\mathrm{B(g.s)}+\alpha$ channel with a small fraction of $^{12}\mathrm{C}(0_{2}^{+})$+p. The yield in the $p_{0}$ from the previous Knudsen data \cite{Knudsen} shows a small very narrow peak at the energy associated with this potential state ($E_p$(lab) = 8.64 MeV) and is taken as $6(2.6)$. The yield in the $p_{1}$ channel is harder to estimate due to the larger background from other states in this region but also shows no evidence of a peak and is also taken to be negligible. Fitting this peak in conjunction with neighboring peaks, the yield in the $\alpha_0$ channel is $18(4.4)$ and yielding $\sigma$ = 280(80) keV and $E_{x} = 11.3(1)$ MeV. In the $p_2$ channel, the yield is $7(2.8)$ with $\sigma$= 220(100) keV and $E_{x}$ = 11.0(1) MeV. These widths are commensurate with the experimental resolution therefore $\Gamma$ is expected to be relatively small ($\Gamma < 200$ keV). This is also corroborated by the small peaks seen in the Knudsen data where the counts appear to lie within one energy bin (width=40 keV) suggesting the width of this state may be even be $\Gamma < 40$ keV but the significance of this peak means this more strict width limit cannot be taken as definite. Given the yields for $\alpha_0$ and $p_2$ are both strong, the spin-parity assignment is favored towards $J^{\pi}=\frac{3}{2}^{-}$ where the angular momentum transfer is L=0 and L=1 respectively. A choice of $J^{\pi}=\frac{1}{2}^{-}$ or $J^{\pi}=\frac{5}{2}^{-}$ would require L=2 for the $\alpha_0$ channel which should heavily suppress the yield and $J^{\pi}=\frac{5}{2}^{-}$ would correspond to L=3 for $p_{2}$ so these options are strongly disfavored.
From Table~\ref{tab:states}, when taking the yield of the states and correcting for the different channel penetrabilities, $P_L$, and efficiencies, one can determine the structure of the measured states without a measurement of the width of the state to compare to the Wigner limit. Many of the states in $^{9}\mathrm{B}$ are very broad and the extreme simplification of calculating the penetrability to the resonant energy is made. In reality, the average penetrability will be higher. The structure is therefore determined by the fractional reduced-width, ${\bar{\gamma}^2_{i}} = \frac{\gamma^2_{i}}{\sum_j \gamma^2_{j}}$ where $\gamma^2_{i} = \frac{\Gamma_{i}}{2P_{iL}}$. This variable shows the type of clustering but not the magnitude of the clustering. This state has considerable strength in both $\alpha_0$ and $p_2$ with ${\bar{\gamma}^2_i}$ as 67\% and 29\% respectively.
Taking the assumption that the total width, $\Gamma$, of the state is $< 200$ keV, one may compare to the Wigner limit, $\gamma_W^2 = \frac{\hbar^2}{\mu a^2}$ which is 0.57 and 2.1 MeV for $\alpha$-decay and $p$-decay respectively. The ratio to the Wigner limit assuming $\Gamma$ = 200 keV is then $\theta_W^2$ $<28$ $\%$ and $<4 \%$ for $\alpha_0$ and $p_2$ respectively. The former of these (while notably only an upper limit) constitutes a well-clustered state.
\subsection{11.8 MeV state}
In the $p_2$ channel, the yield is $4(2.2)$ with $\sigma$= 170(110) keV and $E_{x}$ = 11.8(1) MeV. Counts in the $\alpha_1$ channel are from higher excitation energies extending down as the $P_l$ for $\alpha_1$ is extremely suppressed prohibiting any strength. Due to the strength of the two nearby states in the $\alpha_0$ channel, the yield in the $\alpha_0$ channel has very large uncertainties and can only be limited to be less than $1.8$. There are two states previously known at this energy, a $\frac{5}{2}^{-}$ and a $\frac{3}{2}^{-}$ with widths of 115(30) and 530(80) keV respectively. Our data are more consistent with the $\frac{3}{2}^{-}$ assignment which was ascribed as $\frac{5}{2}^{-}$ in previous work \cite{Knudsen}. A $\frac{5}{2}^{-}$ assignment is the least favored from an angular momentum perspective (L=3 vs L=1 for $\frac{1}{2}^{-}$ or $\frac{3}{2}^{-}$) and this state is seen to populate the $p_2$ channel reasonably well. This ambiguity suggests a reexamination of the total width of each of these two states is needed in future work. From previous work assuming that the strength seen was due to the $\frac{3}{2}^{-}$, the yield in the $p_0$ was determined to be 28$(14)$. Making the same corrections for penetrabilities as above, this state shares strength in the $p_0$ and $p_2$ channels with ${\bar{\gamma}^2_i}$ $>50\%$ and $>38\%$ respectively with the remaining $\alpha_0$ component being $<12\%$. The width for this state is perhaps poorly-known and the reduced width for $p_2$ can be compared to the Wigner limit and is $\sim 1 \%$. Therefore, this state is not strongly $^{12}\mathrm{C}(0_{2}^{+}) \bigotimes p$ clustered (due to the considerable $p_0$ branching ratio).
\subsection{12.4 MeV state}
Fitting this peak in conjunction with neighboring peaks, the yield in the $\alpha_0$ channel is $22(4.8)$, yielding $\sigma$ = 310(90) keV and $E_{x} = 12.4(1)$ MeV. The corresponding yield of $\alpha_{1}$ is $4(2.2)$. In the $p_2$ channel, the yield is $5(2.5)$ with $\sigma$= 110(70) keV and $E_{x}$ = 12.5(1) MeV. Despite the relatively small yield in the $\alpha_1$ channel, when correcting for penetrability, the $\alpha_1$ dominates the strength with ${\bar{\gamma}^{2}_{i}}=88\%$ with $\alpha_0$ and $p_2$ sharing the remainder with 6\% and 2\% respectively. The strong nature of the $^{9}\mathrm{B}(\frac{1}{2}^{+}) \bigotimes \alpha$ suggests this is some kind of near-threshold p-wave state.\par

This energy regime enters the region where existing $^{9}\mathrm{Be}(\alpha,\alpha_0)$ \cite{Be9aa,Ivano} and $^{9}\mathrm{Be}(\alpha,n_0)$ \cite{Obst} are available and one may look for analogous states in $^{13}\mathrm{C}$. Given this state is in the s-wave in the entrance channel (assuming $J^{\pi}=\frac{3}{2}^{-}$) and is expected to be relatively narrow, and previous data seem to have a very large experimental width, it is perhaps possible to explain that such a state has not been observed in this excitation energy in $^{13}\mathrm{C}$ in the $^{9}\mathrm{Be}(\alpha,\alpha_0)$ channel. It can be seen however that the $\alpha_0$ is rather weak and therefore this state may not be strongly populated in this way. The sole dominant feature in this region is a strong $\frac{5}{2}^{+}$ state at 11.95 MeV. \par
It is worth noting that the $\alpha_1$ channel is sub-threshold in $^{13}\mathrm{C}$ and the $n_{2}$ channel is heavily-suppressed until $^{13}\mathrm{C}$ excitation energies of above 13 MeV \cite{Obst}. There are many states in this region ($E_{\alpha} > 2$ MeV) visible in the $^{9}\mathrm{Be}(\alpha,n_0)$ channel but the data are insufficient resolution to provide spin-parity and width assignments.\par

This perhaps motivates a more extensive investigation of near-threshold states in $^{13}\mathrm{C}$ from the $^{9}\mathrm{Be}+\alpha$ channel with higher resolution and angular coverage.
It is also worth noting in the previous proton data \cite{Knudsen} that there is a peak at this corresponding energy for the $p_1$ channel ($E_p$(lab) = 5.55 MeV) where a peak with a yield of $\approx 6$ can be seen above a considerable background. The conservative limit of $<10$ for $p_1$ is therefore taken. The width in this spectrum is also seen to be small which agrees with our results. \par
\subsection{13.1 MeV state}
Around 13.1 MeV, there is a relatively strong component seen in the $\alpha_3$ channel where decays occur through the 2.75 MeV $\frac{5}{2}^{+}$. There is only a very small contribution from the $\alpha_1$ channel at this excitation energy so this state is almost exclusively $^{9}\mathrm{B}(\frac{5}{2}^{+}) \bigotimes \alpha$. Given the dominance of $\alpha_3$, this suggests a spin-parity of $J^{\pi} = \frac{5}{2}^{-}$ which suppresses the other channels. \par

In $^{9}\mathrm{B}$, there is also the extremely-broad 2.78 MeV $\frac{1}{2}^{-}$ with $\Gamma$ = 3.13 MeV which cannot be excluded as the source of the $\alpha_3$ strength. The reason for this is because we are decaying into a broad intermediate state, the penetrability to lower excitation energies in $^{9}$B is exponentially enhanced therefore meaning the width-dependent Breit-Wigner shape is hugely distorted towards lower excitation energies. Simple R-Matrix calculations show that the $\frac{1}{2}^{-}$ being fed by an excitation energy of around 13 MeV in $^{13}$N with L=0 may produce a yield in the $^{9}$B relative energy space that looks similar to the narrower 2.75 MeV $\frac{5}{2}^{+}$ in $^9$B. Our data do not have sufficient statistics to exclude this possibility and the $\frac{1}{2}^{-}$ decays primarily through $^{8}\mathrm{Be}$ via proton-decay. In this possibility, the preferred spin-parity assignment is obviously $J^{\pi} =\frac{1}{2}^{-}$ corresponding to L=0 $\alpha_3$ decay. The results for both spin parities assignments are included in Table~\ref{tab:states}.\par
As with the 12.4 MeV state, there is evidence of a peak in previous data at the correct energy in the $p_1$ channel ($E_p$(lab) = 6.20 MeV) which is given a similar limit of $<10$.
\subsection{13.7 MeV state}
There is a collection of strength in the $p_2$, $\alpha_0$, $\alpha_1$ and $\alpha_3$ channel. With a yield of $6(2.7)$, the state is dominated by $p_2$ and has parameters of $\sigma$= 260(70) keV and $E_{x}$ = 13.7(1) MeV. Given the large $\bar{\gamma^2}$ in the $\alpha_3$ channel, this state can be assigned as either $\frac{3}{2}^{-}$ or $\frac{5}{2}^{-}$. A $\frac{5}{2}^{-}$ would correspond to L=3 for the $p_2$ channel so a $\frac{3}{2}^{-}$ assignment would be more commensurate with the reasonable $p_2$ yield. This state also exhibits a $^{9}\mathrm{B}(\frac{5}{2}^{+}) \bigotimes \alpha$ structure. \par
Examining the previous work for evidence of a peak in the $p_1$ is not possible for this state due to the presence of a strong $p_0$ branch from a lower-lying state at the same energy. A similar limit of $<10$ is therefore placed on this state.
\begin{figure}
\centerline{\includegraphics[width=0.5\textwidth]{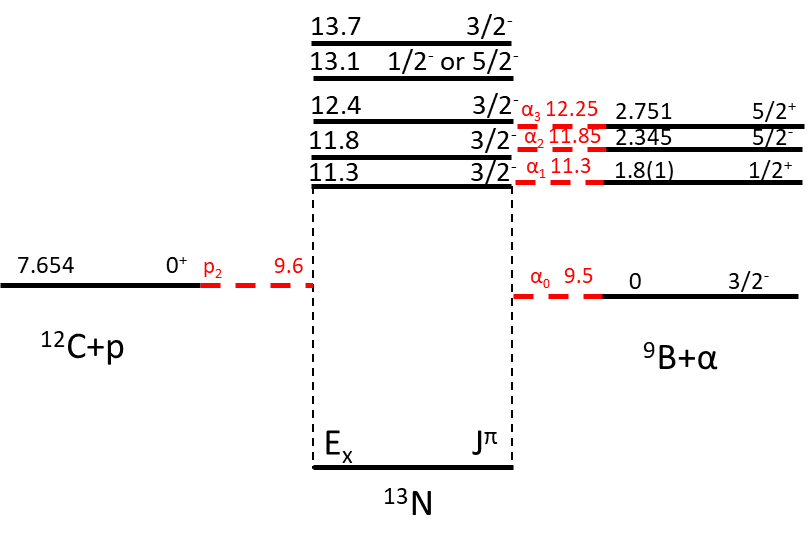}}
\caption{Level scheme of measured 3$\alpha$+p states in $^{13}\mathrm{N}$ in the central column with the proposed spin-parity assignments. The location of the thresholds for proton and $\alpha$ decay are shown in red with the equivalent excitation energy shown. The corresponding states in the daughter nuclei ($^{12}\mathrm{C}$ and $^{9}\mathrm{B}$) are similarly displayed. Also shown in \cite{PRL}. \label{fig:levelscheme}}
\end{figure}
\section{Conclusions}
Three new states and a previously-tentative state have been observed with a strong $3\alpha+p$ nature with their excitation energies relative to the thresholds shown in Fig.~\ref{fig:levelscheme}. The first is a narrow potential $\frac{3}{2}^{-}$ state at $E_{x}$ = 11.3(1) MeV with mixed $^{9}\mathrm{B}(\mathrm{g.s}) \bigotimes \alpha$ and $p+^{12}\mathrm{C}(0_{2}^{+})$ nature. \par
Another previously-observed $\frac{3}{2}^{-}$ was seen to have mixed $p+^{12}\mathrm{C}(\mathrm{g.s.})$ and $p+^{12}\mathrm{C}(0_{2}^{+})$ nature at 11.8 MeV with around half of the total strength as $p+^{12}\mathrm{C}(\mathrm{g.s.})$. \par
At higher excitation, another strong $\alpha$-decaying state was seen at $E_{x}$ = 12.4(1) MeV although this state has a much stronger $^{9}\mathrm{B}({\frac{1}{2}^{+}}) \bigotimes \alpha$ nature. \par
A revised excitation energy of 13.1(1) MeV is suggested for a previously-seen state at 13.26 MeV. The $^{9}\mathrm{B}({\frac{5}{2}^{+}}) \bigotimes \alpha$ structure dominates in this state and a spin assignment of $J^{\pi}$ = ${\frac{1}{2}^{-}}$ or ${\frac{5}{2}^{-}}$ are therefore preferred. \par
Finally, another $\frac{3}{2}^{-}$ is seen at 13.7 MeV which is also dominated by $^{9}\mathrm{B}({\frac{5}{2}^{+}}) \bigotimes \alpha$. \par
The inability to extract the width of these narrow states means that the magnitude of clustering cannot be fully evaluated however the type (channel) of clustering can be determined without this information. Higher resolution data focusing on the proton channel may provide further information on the magnitude of this clustering phenomenon. From our current data, one may conclude however when comparing the reduced widths that the clustered channels are very competitive against the single-particle $p_0$ channel. Configuration mixing may however be very strong in these states and therefore quench the single-cluster nature of these resonances. \par

One can compare these states with $3\alpha$+p events observed via other reactions such as those populated by single proton decay from highly-excited states in $^{14}$O$^{\star}$ in complete kinematics \cite{Charity}. Due to the spin-parity selectivity of $\beta$-decay and the unusual population method of states in $^{13}$N$^{\star}$ in the previous work, it is perhaps not surprising that the same states are not observed in the current work.\par
Work using $^{13}$C($^{3}$He,t) by Fujimura et al. has also been performed which, at high energy, is a better analogue to inverse $\beta$-decay and should preferentially populate similar states although the ground state spin-parity of $^{13}$C is $J^{\pi}=\frac{1}{2}^{-}$ rather than $J^{\pi}=\frac{3}{2}^{-}$ as for $^{13}$O \cite{Fujimura}. This work relies on the excitation energy being extracted by missing mass and the decay channels tagged solely by protons to separate $p_0$, $p_1$, $p_2$ etc. A broad peak at $E_x$ = 13.5 MeV can be seen in the $p_2$ channel. This lies close in energy to our observed peak at 13.8 MeV which, with a preferred $J^{\pi}=\frac{3}{2}^{-}$ agrees with the state being observed with ($^{3}$He,t). The 11.8 MeV states, also observed previously by Knudsen was also seen to be very strongly populated in the Fujimura data. Their data show a smaller $p_0$ and $p_2$ yield and a dominant $p_1$ yield which is at odds with the Knudsen result which ascribed the yield in this region to the $J^{\pi}=\frac{5}{2}^{-}$ which would not be populated strongly by Fujimara. Is it therefore possible that in the current work and that of Knudsen, the $J^{\pi}=\frac{3}{2}^{-}$ and $J^{\pi}=\frac{5}{2}^{-}$ are both contributing strength here in different channels. Alternatively, in the work by Knudsen the $p_1$ yield could have been partially obscured by the strong $p_0$ yield for the 7.376 MeV state. Further studies with higher spin-sensitivity are necessary to disentangle these two contributions however.\par
We hope this experimental work will motivate further theoretical studies for $^{13}$N including using the Algebraic Cluster Model (D$^{'}_{3h}$) as performed for $^{13}$C \cite{d3h} and AMD calculations.
\section{Acknowledgments}
This work was supported by the U.S. Department of Energy, Office of Science, Office of Nuclear Science under Award No. DE-FG02-93ER40773 and by the National Nuclear Security Administration through the Center for Excellence in Nuclear Training and University Based
Research (CENTAUR) under Grant No. DE-NA0003841. G.V.R. also acknowledges the support of the Nuclear Solutions Institute. S.A., S.M.C., C.K., D.K., S.K. and C.P. also acknowledge travel support from the IBS grant, funded by the Korean Government under grant number IBS-R031-D1. C.H.K acknowledges travel support from the National Research Foundation of Korea(NRF) grant, funded by the Korea government(MSIT) (No. 2020R1A2C1005981 and 2013M7A1A1075764).
\bibliography{PRCBib}

\end{document}